\documentclass[a4paper]{jpconf}
\usepackage{graphicx}
\usepackage{subcaption}

\usepackage{soul}

\usepackage{url}
\usepackage{xcolor}
\usepackage{hyperref}
\usepackage{amsmath}
\usepackage{verbatimbox}
\usepackage{float} 
\usepackage{lineno}

\usepackage{fancyhdr}
\renewcommand{\sc}{\textsc}
\renewcommand{\bf}{\textbf}
\renewcommand{\it}{\textit}
\renewcommand{\tt}{\texttt}

\def\MJ{\sc{Majorana}}
\def\DEM{\sc{Demonstrator}}
\def\MJD{\sc{Majorana Demonstrator}}

\begin{document}
\title{Beyond the Standard Model Searches with the \MJD\ Experiment}

\author{Wenqin Xu for the \MJ\ Collaboration}

\address{University of South Dakota, Vermillion, SD, 57069, USA}

\ead{Wenqin.Xu@usd.edu}

\begin{abstract}
The \MJD\ experiment operated two modular arrays of p-type point contact high purity germanium (HPGe) detectors, of which 30 kg is enriched to 88\% in Ge-76, to search for neutrinoless double beta decay. The data-taking campaign for double beta decay with enriched detectors was successfully concluded in March 2021, and data-taking with natural detectors is still ongoing. The \DEM~has achieved excellent energy performance in a wide dynamic range covering 1 keV to 10 MeV. The extra-low background level and excellent energy performance achieved by the \DEM~ makes it competitive in various searches of physics beyond the Standard Model. 

If there is an axion-photon coupling, axions can be produced by the Primakoff conversion of photons in the Sun. Solar axions can inversely generate photon signals in germanium crystals, which can be coherently enhanced when the Bragg condition is satisfied. The \DEM~is searching for solar axions with a novel method to correlate and leverage its high number of HPGe detectors. We will discuss the status and results of recent searches for new physics with the \DEM, including the first reporting of a solar axion search.
\end{abstract}
\section{Introduction}
The \MJD\ experiment~\cite{MJD2014} searches for neutrinoless double-beta decay ($0\nu\beta\beta$) of $^{76}$Ge in HPGe detectors as well as additional physics Beyond the Standard Model (BSM). Low background was accomplished by using ultra-clean materials and deploying HPGe detectors within a compact graded shield with an active muon veto at the 4850 foot level of the Sanford Underground Research Facility (SURF). An excellent energy resolution approaching 0.1\% full-width-at-half-maximum at the Q-value of $^{76}$Ge $0\nu\beta\beta$ decay has been achieved. Recent $0\nu\beta\beta$ results from the \DEM\ can be found in Ref.~\cite{MJD2019,MJD2021}. With limited cosmogenic exposure on the surface, \MJ\ enriched detectors have ultra-low backgrounds at low energy, for example about 0.01 cts/kg-d/keV between 20–40 keV~\cite{WisemanTAUP2020}. Excellent energy performance and ultra-low background allows the \DEM\ to perform multiple competitive BSM searches, \it{e.g.}~\cite{MJD2017, MJD2018, MJD2019PRD}.~
Axions and pseudoscalar axion-like particles (ALPs) are candidates for low-mass dark matter. Recent \MJ\ search results on axions and ALPs are described in these proceedings.
\section{Pseudoscalar ALP Search}
\begin{figure}[!htb]
    \centering
    \begin{subfigure}[t]{0.55\textwidth}
    \centering
    \includegraphics[width=\textwidth]{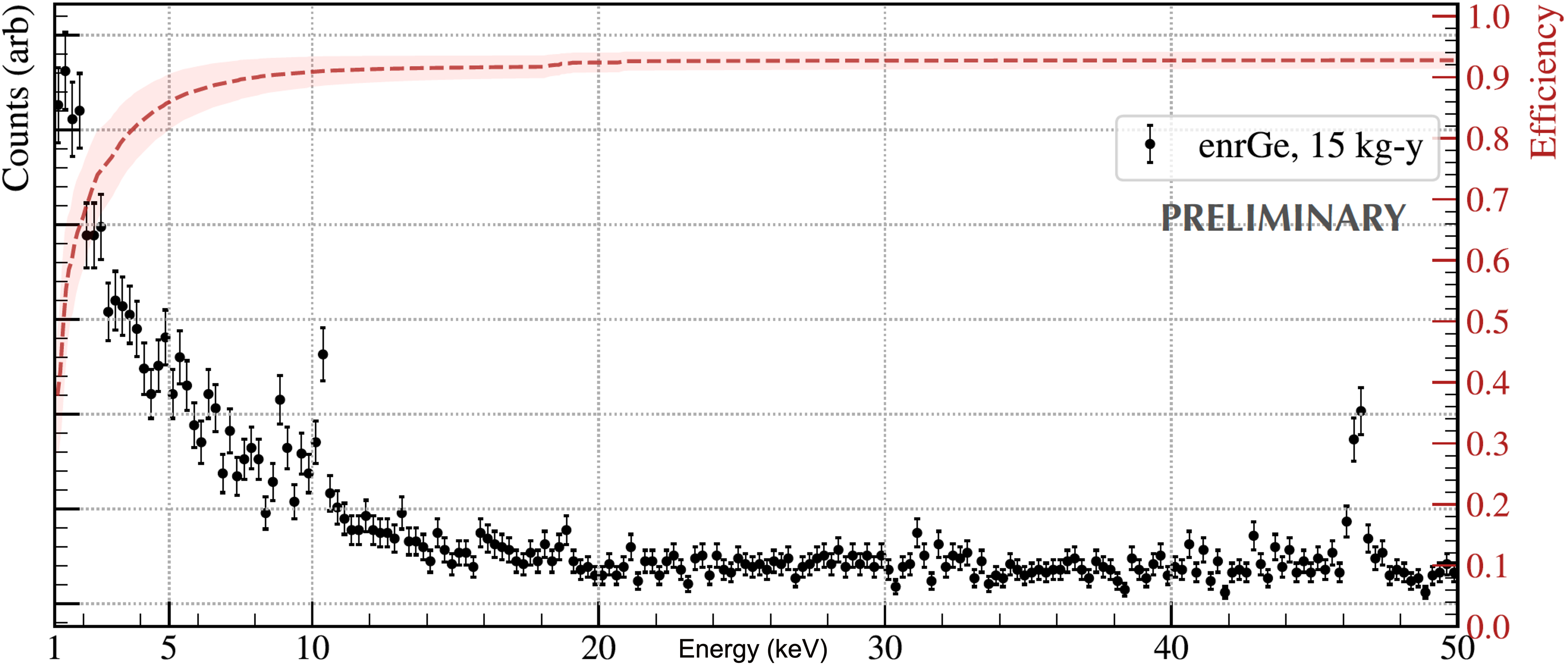} 
    \end{subfigure}
    \begin{subfigure}[t]{0.4\textwidth}
    \centering
    \includegraphics[width=\textwidth]{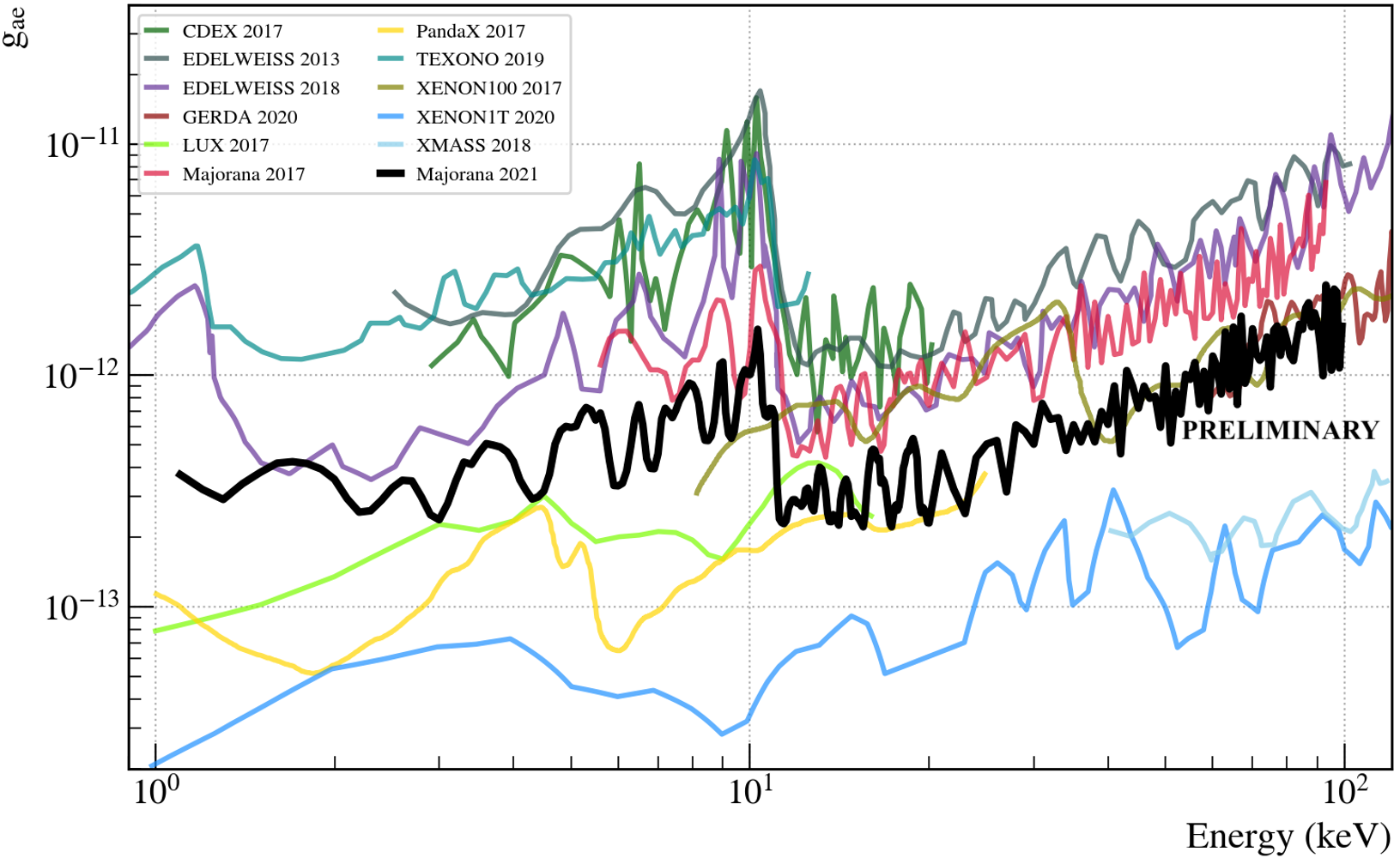} 
    \end{subfigure}
    \caption{(Adapted from~\cite{WisemanAPS2021}) Left: Preliminary energy spectrum (black data points) and combined efficiency (red dashed curve) for enriched detectors in open data set in DS1 through DS6C. Right: Preliminary exclusion limits (black curve) on axion-electron coupling from this DS1–DS6C analysis, plotted along with other limits (see legend).}
    \label{fig:APS2021}
\end{figure}

\MJ\ reported its first search result of ALPs in the dark matter halo in~\cite{MJD2017}. As presented at the last TAUP conference (TAUP2019)~\cite{WisemanTAUP2020}, a pulse-shape based novel method for slow pulse determination was developed and an analysis using about 11.2 kg-yr exposure resulted in improved axion-electron coupling limits for ALPs between 5 and 100 keV. The efficiency calculation was based on low-energy, small-angle Compton scatter events in the calibration data~\cite{WisemanTAUP2020}. This slow pulse cut and a range of techniques with careful pulse shape analysis and threshold optimization are applied in the latest analysis of 15 kg-yr open data in dataset (DS) 1 through 6C. The noise at the lowest energies has been reduced by about 5 orders of magnitude and the analysis threshold has been lowered from 5 keV to 1 keV, as shown in the left panel of Fig.~\ref{fig:APS2021}. A preliminary ALP analysis similar to the 2017~\cite{MJD2017} and 2019~\cite{WisemanTAUP2020} analyses generated new axion-electron coupling limits from the \DEM\, as shown in the right panel of Fig.~\ref{fig:APS2021}. Blind data are still to be analyzed, which will roughly double the exposure.

\section{Solar Axion Search}
The sun could be a major source of axions. Assuming only axion-photon coupling, thermal photons in the electromagnetic field of the solar plasma can become axions via inverse Primakoff conversion~\cite{Raffelt1988}, with energies mostly below 10 keV. One recent parameterization of solar axion flux resulting from Primakoff production is given in Ref.~\cite{Raffelt2008, Arik2011}. For a recent review of solar axion detection, see Ref.~\cite{Irastorza2018}. Solar axions entering a solid state detector can be converted back into photons in the intense atomic electric field via Primakoff conversion. When the Bragg condition is met, conversions from multiple crystal planes will be coherently enhanced in a fashion similar to x-ray diffraction, generating strong solar axion signals with a distinct pattern~\cite{Buchmuller1990, Paschos1994, Creswick1998, Cebrian1999}. This formalism of coherent Primakoff-Bragg conversion is valid for axion mass up to a few hundred eV, and it has been explored by previous solid state experiments, including SOLAX~\cite{Avignone1998}, DAMA~\cite{Bernabei2001}, COSME~\cite{Morales2002}, CDMS~\cite{Ahmed2009}, and EDELWEISS II~\cite{Armengaud2013}. The current best limit with 90\% Confidence Level (CL) is g$_{a\gamma\gamma}<1.7\times 10^{-9}$ GeV$^{-1}$ from DAMA and the best 95\% CL limit is g$_{a\gamma\gamma}<2.15\times 10^{-9}$ GeV$^{-1}$ from EDELWEISS II. These limits are better than current helioscope limits for axion mass above roughly 1 keV.

\begin{figure}[!htb]
    \centering
    \begin{subfigure}[t]{0.57\textwidth}
    \centering
    \includegraphics[width=\textwidth]{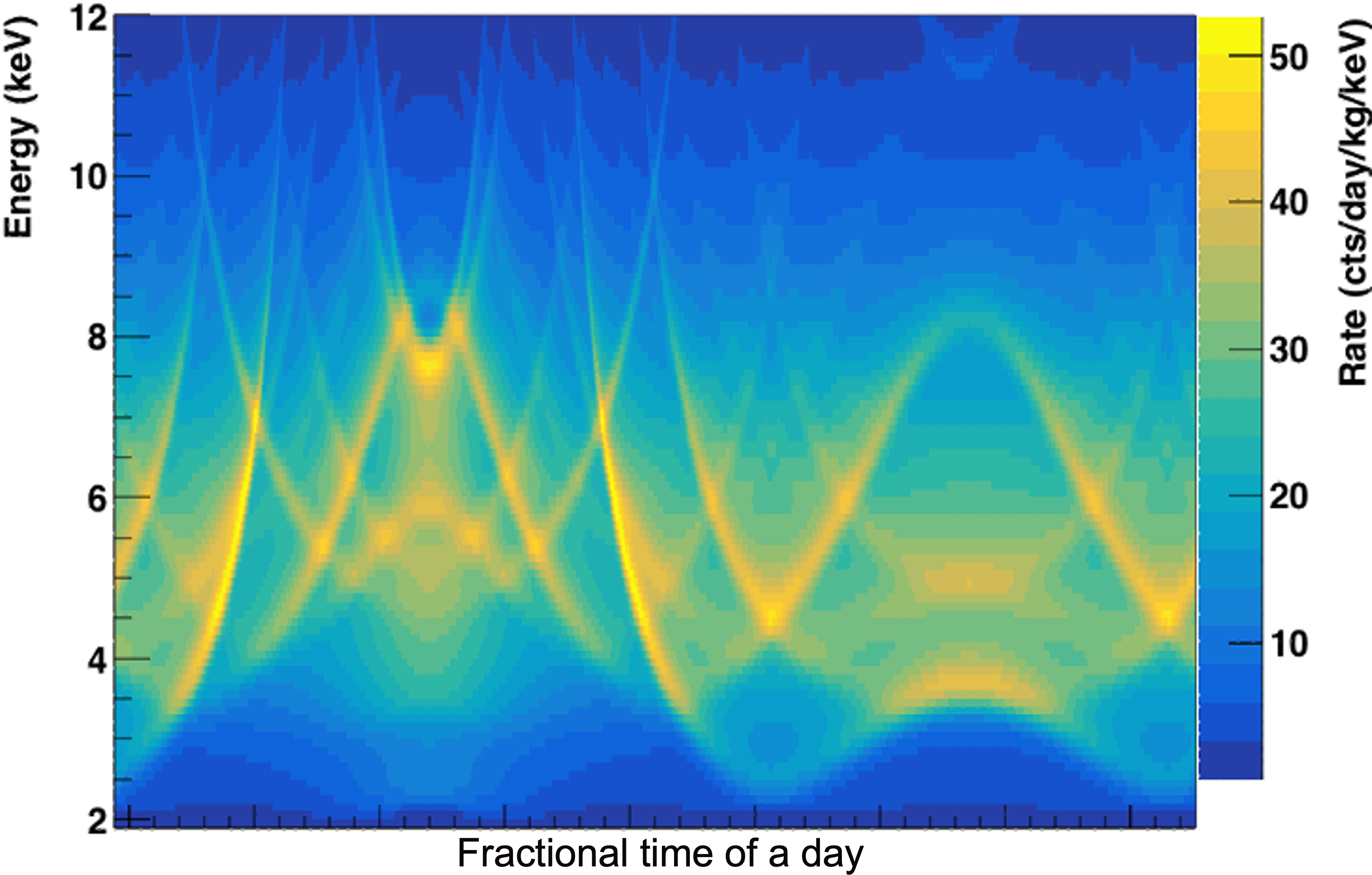}
    \end{subfigure}
    \begin{subfigure}[t]{0.42\textwidth}
    \centering
    \includegraphics[width=\textwidth]{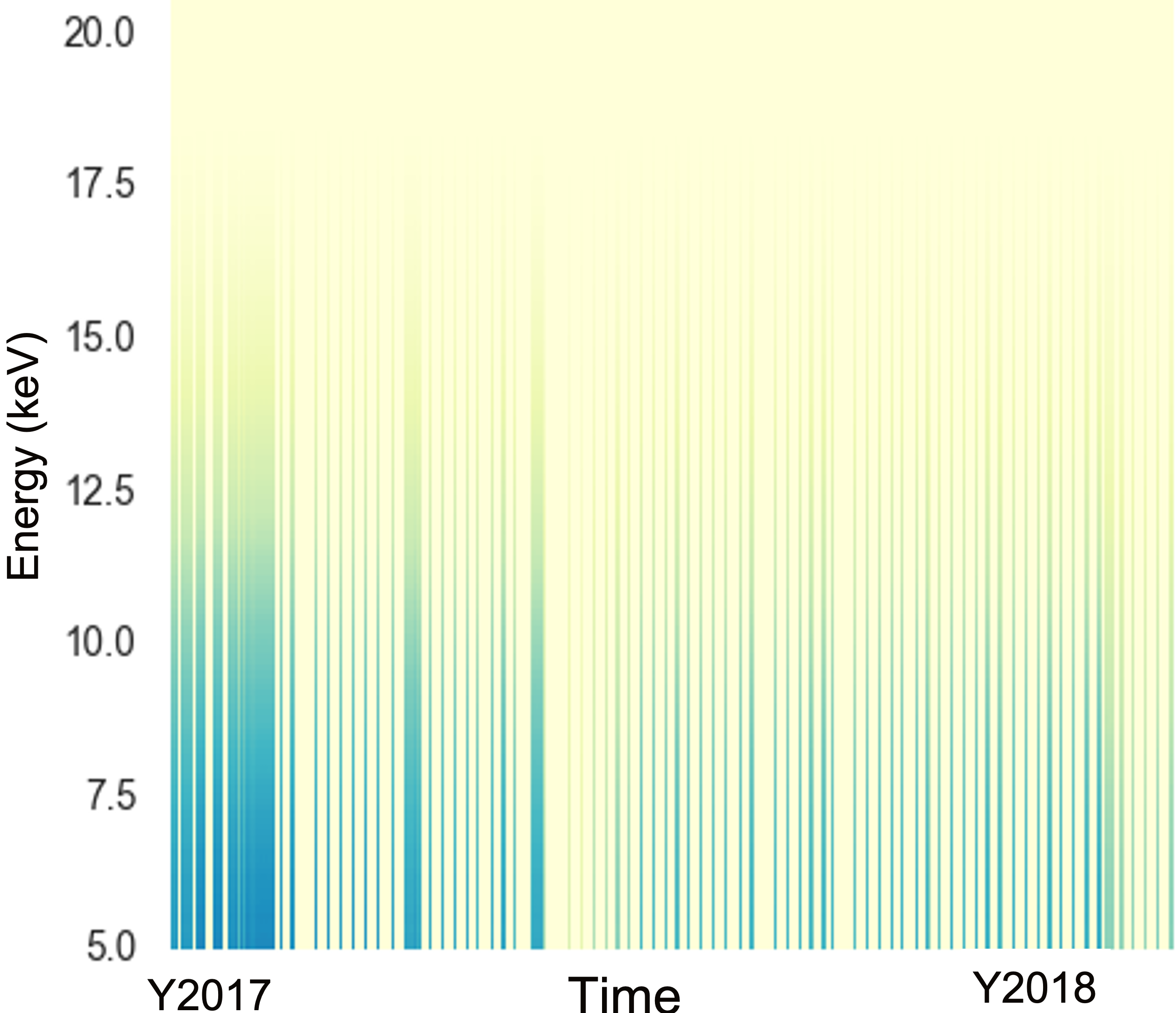}
    \end{subfigure}
    \caption{Left: The angle-averaged solar axion signal as a function of both time and energy for a HPGe detector located at SURF, assuming g$_{a\gamma\gamma}=1\times 10^{-8}$ GeV$^{-1}$. Right: The 2-dimensional tritium PDF. The displayed time range is 1 day for axion and about 1 year for tritium. Data-taking duty cycles have been folded into the tritium PDF.}
    \label{fig:axion_pdf}
\end{figure}

Predicting the exact probability distribution function (PDF) of solar axion signals in a Ge crystal requires knowledge on both the solar position, which comes from United States Naval Observatory database~\cite{USNO} in this analysis, and the orientations of three main crystallographic axes, which are perpendicular to each other. While the manufacturing process of HPGe detectors makes one axis vertical, the orientations of the two axes on the horizontal plane are unknown without further measurements and tracking during installation. So far, only CDMS measured the horizontal axis angle in one deployed detector~\cite{Ahmed2009}. The current best 90\% CL limit from DAMA~\cite{Bernabei2001} was obtained by averaging the solar axion PDF over all possible angles of the horizontal lattices in their NaI crystals. This angle-averaging technique was carefully examined in Ref.~\cite{Xu2017} in the Frequentist framework and justified for experiments with at least 15 detectors. In the Bayesian framework, the horizontal angle is a nuisance parameter and it can be marginalized over. The prior distribution of the angle is evidently uniform given the completely random deployment of crystals. Marginalizing over the angle with a flat prior leads to the angle-averaged PDF, which is shown in the left panel of Fig.~\ref{fig:axion_pdf}. 

\begin{figure}[!htb]
    \centering
    \includegraphics[width=0.95\textwidth]{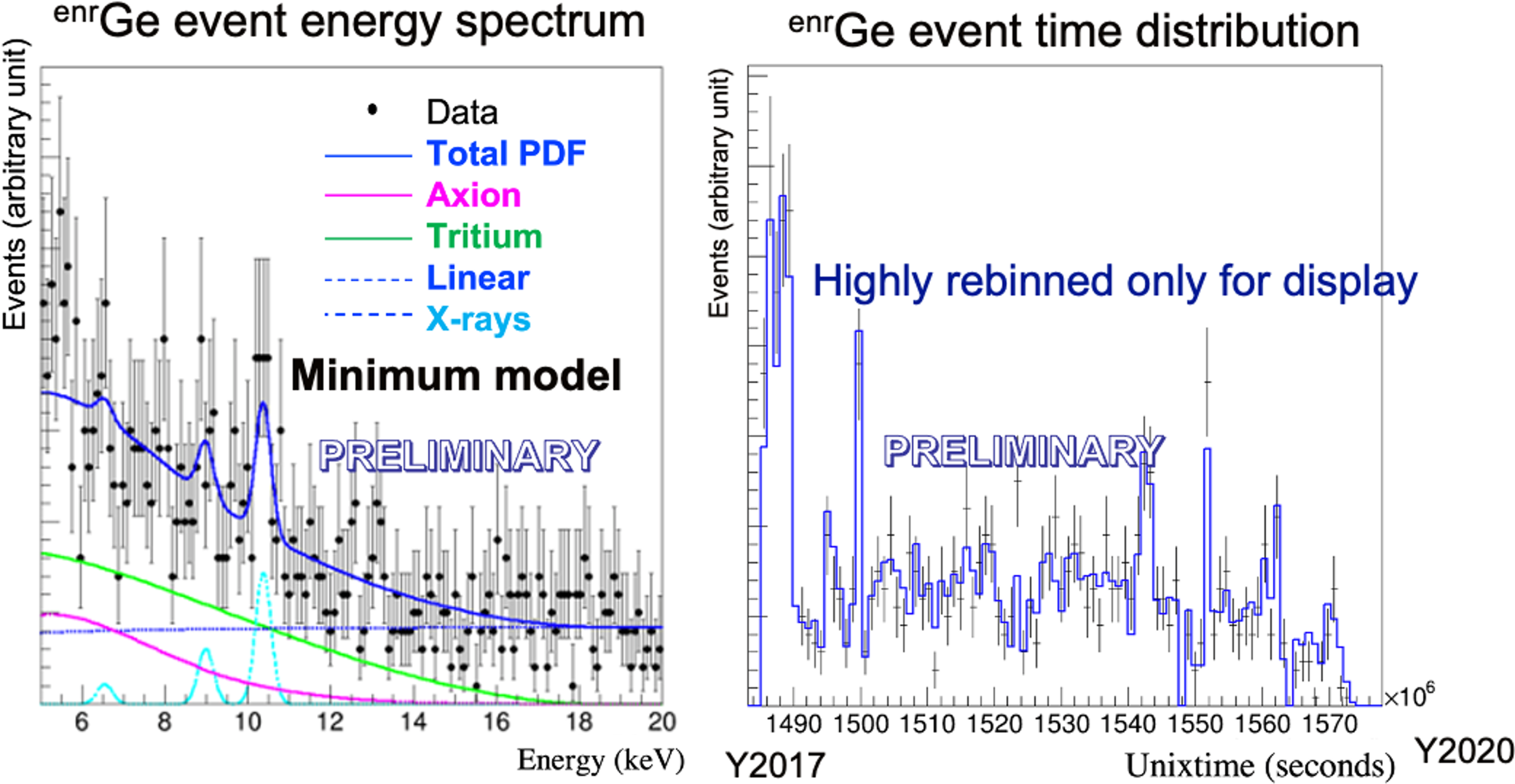}
    \caption{Data and best fit of the minimum model in the energy (left) and temporal dimensions (right). In both plots, the black points are data and solid blue curves are the best fit composite model, projected into one dimension.}
    \label{fig:fits}
\end{figure}

The solar axion search is carried out using about 10 kg-yr open data in DS5B through 6C, in which the noise level was low and at least 20 enriched detectors can be used in the analysis. A conservative background modeling, called the minimum model, is considered here, including the tritium $\beta$-decay continuum, a linear background representing an extrapolation of the Compton continuum, and three x-ray peaks from $^{55}$Fe (6.54 keV), $^{65}$Zn (8.98 keV), and $^{68}$Ge (10.37 keV). All components in the minimum model are expected to be present, even if at minimal levels. To fully exploit the distinctive axion signature, the analysis is performed in both temporal and energy dimensions simultaneously. The axion and background PDFs are folded with energy-dependent efficiency as well as data-taking duty cycles as a function of time. Blind data-taking periods are considered as periods without data for this open data analysis. While tritium and x-ray backgrounds are modelled to decay away with their half-lives (\it{e.g.} see the right panel of Fig.~\ref{fig:axion_pdf} for tritium), the Compton scattering term is assumed constant in time. The efficiency-folded axion signal spectrum peaks at around 5-6 keV and reaches almost zero beyond 12 keV. The energy range of the current analysis is chosen to be between 5 and 20 keV.

\begin{figure}[!htb]
    \centering
    \includegraphics[width=0.8\textwidth]{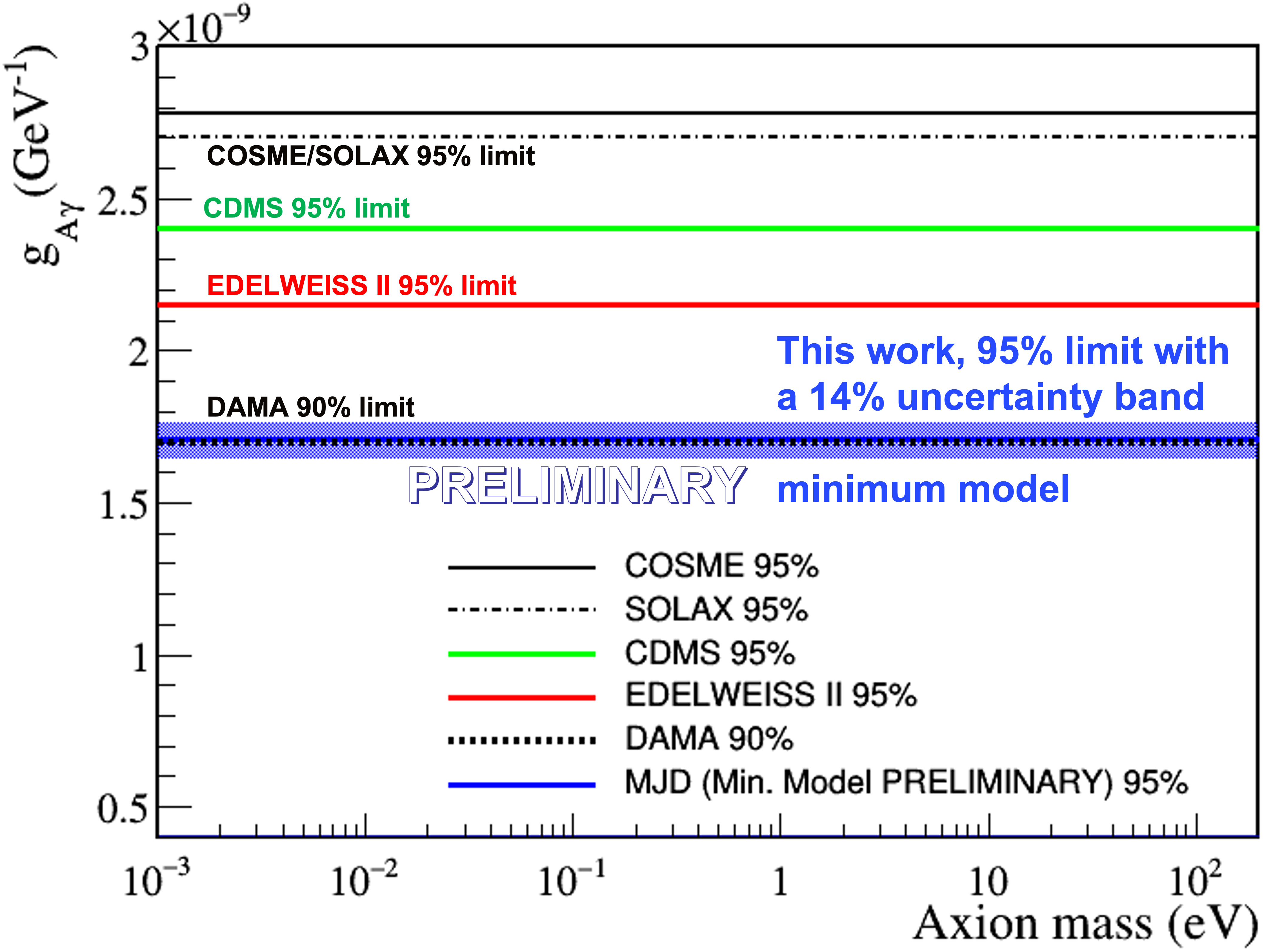}    
    \caption{Axion-photo coupling limits from this analysis (solid blue line) and other solid-state experiments (see legend).}
    \label{fig:limits}
\end{figure}

To display both the data and the composite model of signal and background, a maximum likelihood best fit of the model to the data is shown in Fig.~\ref{fig:fits}. The only unknown parameters are the strengths for all components. To obtain a Bayesian limit on axions, posteriors of unknown parameters are sampled by the Markov-Chain Monte Carlo (MCMC) method, utilizing the Metropolis-Hastings (MH) sampling algorithm implemented in \texttt{RooStats}~\cite{RooStats}. A uniform prior between zero and the total number of events in the data is used for axions. A 95\% credible interval (CI) starting from zero is obtained, which translates to a preliminary 95\% CI limit of $g_{a\gamma\gamma}<1.71\times 10^{-9}$ GeV$^{-1}$. This is better than the previous 95\% limits from solid state experiments and comparable to the current best 90\% limit, as shown in Fig.~\ref{fig:limits}. Major uncertainties considered so far include the combined analysis cut efficiency (8\%), the energy determination (5\%), and a different software (PyMC3~\cite{PYMC3}) using No-U-Turn (NUTS) sampling algorithm (10\%). In total, the systematic uncertainty on the preliminary limit is 14\%.

\section{Summary}
In summary, by further reducing noise and lowering the analysis threshold at the \DEM, new axion-electron coupling limits are obtained for ALPs between 1 and 100 keV, improving and extending previous \MJ\ limits. A solar axion search leads to a preliminary 95\% CI limit on the axion-photon coupling, which is better than previous 95\% limits from solid-state experiments. These open data analyses will be improved when including the blind data. 

\section{Acknowledgements}
This material is based upon work supported by the U.S. Department of Energy, Office of Science, Office of Nuclear Physics, the Particle Astrophysics and Nuclear Physics Programs of the National Science Foundation, the Russian Foundation for Basic Research, the Natural Sciences and Engineering Research Council of Canada, the Canada Foundation for Innovation John R. Evans Leaders Fund, the National Energy Research Scientific Computing Center, and the Oak Ridge Leadership Computing Facility, and the Sanford Underground Research Facility.

\end{document}